\documentclass[reprint,aps,prx,superscriptaddress,amsmath,amssymb,floatfix,footinbib,longbibliography]{revtex4-1}
\usepackage{lineno}
\usepackage{xcolor}
\usepackage[normalem]{ulem}
\usepackage{bm}
\usepackage[colorlinks, linkcolor= blue, citecolor = blue, urlcolor=blue]{hyperref}
\usepackage{physics}
\usepackage{float}
\usepackage{wrapfig}
\usepackage[stable]{footmisc}
\usepackage{array}
\usepackage{multirow}
\usepackage{gensymb}
\usepackage{siunitx}
\usepackage{hhline}
\usepackage{pifont}
\usepackage{chemformula}
\usepackage[overload]{textcase} 
\usepackage{multirow}
\usepackage{tabularx}
\usepackage{array}

\def\nn{\nonumber}
\def\bea{\begin{eqnarray}}
\def\eea{\end{eqnarray}}
\def\be{\begin{equation}}
\def\ee{\end{equation}}

\def\bal{\begin{aligned}}
\def\eal{\end{aligned}}

\usepackage{babel}

\begin{document}

\title{Nonlinear Spin Polarization Enables N\'eel Order Switching in Centrosymmetric Altermagnets}

\author{Sunit Das}
\email{sunitd@iitk.ac.in}
\thanks{Joint first author and equal contribution.}

\author{Sayan Sarkar}
\email{sayans21@iitk.ac.in}
\thanks{Joint first author and equal contribution.}

\author{Amit Agarwal}
\email{amitag@iitk.ac.in}
\affiliation{Department of Physics, Indian Institute of Technology Kanpur, Kanpur 208016, India}

\begin{abstract}
   Deterministic electrical switching of the N\'eel order remains challenging in centrosymmetric altermagnets, where linear current-induced spin polarization is forbidden by inversion symmetry. Here, we establish a nonlinear route to N\'eel order control. Spin group symmetry shows that, in the nonrelativistic limit, the nonlinear spin polarization is constrained along the N\'eel vector, with a staggered response allowed in $8$ of the $10$ nontrivial spin-Laue groups. Although this N\'eel spin polarization is torque-inactive by itself, finite spin-orbit coupling can generate an additional uniform component transverse to the N\'eel vector. Using a minimal model for the $d$-wave altermagnet \ch{FeSb2}, we demonstrate the coexistence of the nonrelativistic staggered response with an SOC-induced uniform transverse component. Landau-Lifshitz-Gilbert equation-based macro-spin simulations reveal that the transverse component alone produces oscillatory dynamics, whereas its cooperative action with the staggered response enables deterministic $180^\circ$ reversal. We further show that changing the electric field orientation reverses the staggered spin response and selects the switching direction, while the opposite N\'eel states can be distinguished by their anomalous Hall response. Our results establish a nonlinear electrical write-read scheme for centrosymmetric altermagnets.
\end{abstract}

\maketitle

\section{Introduction} Altermagnets have recently emerged as a distinct class of magnetically ordered materials that combine compensated magnetization with pronounced nonrelativistic spin-split electronic band structures~\cite{smejkal_prx2022, Smejkal_prx22_emerging, jungwirth_arxiv2024, Song_NRM2025, Jungwirth_nat26}. Unlike conventional antiferromagnets, this spin splitting can persist even in the absence of relativistic spin–orbit coupling, enabling ferromagnet-like transport responses while retaining zero net magnetization. Importantly, many altermagnets exhibit a sizable anomalous Hall effect~\cite{Gonzalez_prl23, Attias_prb2024, Reichlova_NC2024, Sato_prl2024, Tschirner_APL2023}, providing a direct electrical readout of the N\'eel vector orientation, thereby overcoming a key bottleneck in antiferromagnetic memory technologies~\cite{Zhang_prb2022,wang_prl2023,kimel_JOM2024}. Together with negligible stray fields, robustness against external magnetic perturbations, and ultrafast N\'eel order dynamics, these features position altermagnets as promising platforms for designing ultrafast, energy-efficient spintronic and memory applications~\cite{Hayami_JPSJ2019,Hayami_prb2020, Rathore_JAC2025, Tamang_PP2025, Fukaya_JPCM2025,Urru_prb2025}.

\begin{figure}[t]
    \centering
    \includegraphics[width=1\linewidth]{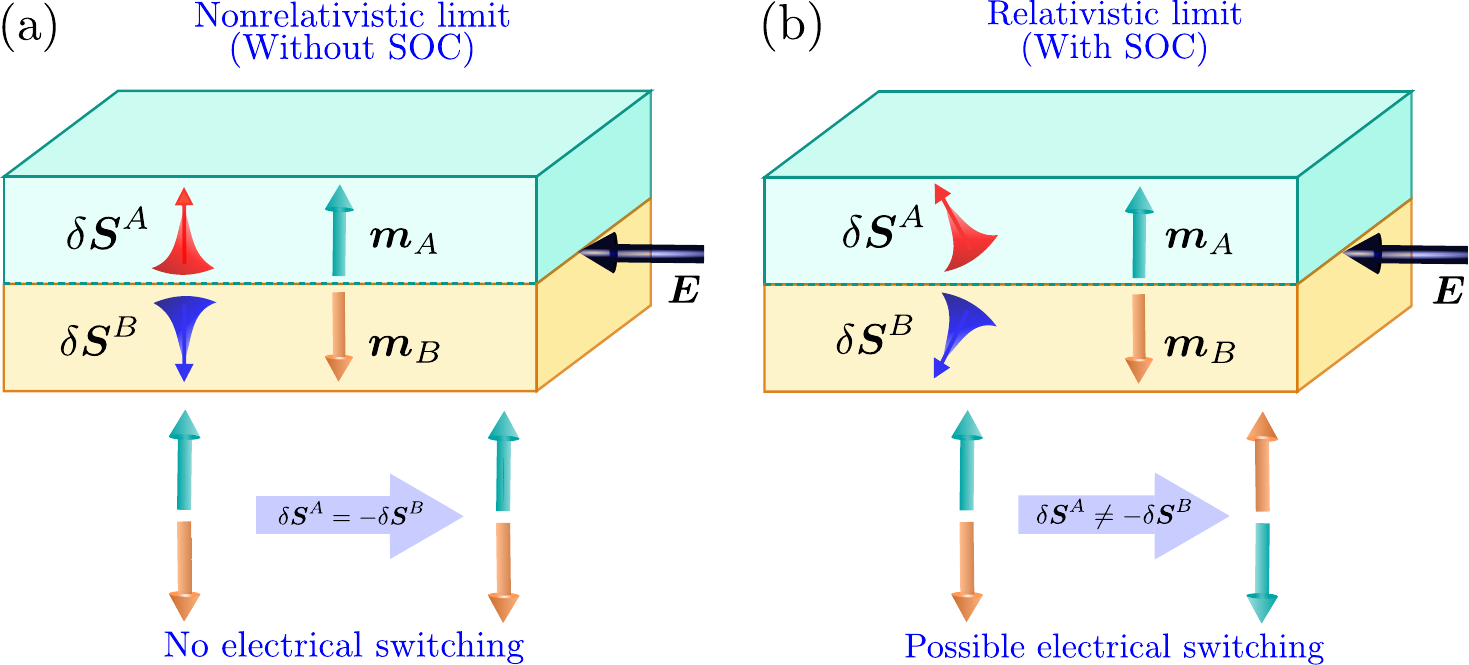}
    \caption{\textbf{Schematic of nonlinear spin polarization-driven N\'eel order switching.}
    (a) In the nonrelativistic limit, centrosymmetric altermagnets can host a staggered nonlinear spin polarization collinear with the N\'eel vector, $\delta\bm S^{A}=-\delta\bm S^{B}$. Being parallel to the local moments, this response is torque-inactive.
    (b) Finite SOC generates an additional transverse spin component, so that the total sublattice spin polarizations are no longer exactly opposite, $\delta\bm S^{A}\neq-\delta\bm S^{B}$. The resulting coexistence of transverse and staggered collinear spin polarization components produces the torques required for electrical reversal of the N\'eel order.}
    \label{fig1}
\end{figure}

However, a central challenge in altermagnetic spintronics is the deterministic electrical control of the N\'eel order. In altermagnets with broken inversion symmetry, a linear current-induced spin polarization can generate spin-orbit torque and enable N\'eel vector switching~\cite{Wadley_nature24,Chen_prl2025,Zhou_NC2025, Sarkar_26_asym}. Such a linear response is, however, forbidden in the large class of even-parity altermagnets that preserve inversion symmetry~\cite{Edelstein1990, SARKAR2026100062, jo2026}. Several alternative routes have been explored, including spin-splitter effects, magnetic-octupole injection, asymmetric sublattice spin currents, and asymmetric spin torque~\cite{Zelenzy_prl2021,Han_25,Lee_prb25,sarkar2025,Tsong_prb25,Yang_neel25, Shao_25}. Despite these advances, identifying a broadly applicable mechanism for deterministic, all-electrical switching in centrosymmetric altermagnets remains an important challenge.

In this work, we establish a distinct switching mechanism arising from the interplay of nonrelativistic altermagnetic spin polarization and relativistic spin-orbit coupling (SOC). Using spin group symmetry, we first show that in the nonrelativistic limit the nonlinear spin polarization is constrained to lie along the N\'eel vector and can have a staggered component between the two magnetic sublattices. Such nonlinear N\'eel spin polarization is symmetry allowed in 8 of 10 nontrivial spin-Laue groups, but its collinearity with the local moments renders it torque-inactive. Finite SOC qualitatively changes this situation by allowing an additional transverse nonlinear spin polarization. As we demonstrate below, the coexistence of the collinear N\'eel and transverse spin responses provides the essential ingredients for deterministic electrical reversal of the N\'eel order in centrosymmetric altermagnets.

Using a minimal model for the $d$-wave altermagnet \ch{FeSb2}, we explicitly demonstrate the coexistence of a nonrelativistic staggered spin polarization parallel to the N\'eel vector and an SOC-induced uniform transverse spin polarization. The resulting nonequilibrium spin polarizations on the two magnetic sublattices are consequently no longer exactly opposite, providing the microscopic origin of the nonlinear sublattice torque responsible for the N\'eel order dynamics. We evaluate these sublattice torques and incorporate them into macro-spin simulations based on the Landau-Lifshitz-Gilbert equation, demonstrating deterministic $180^\circ$ electrical switching of the N\'eel vector. Our results thus establish a distinct mechanism for all-electrical control of both locally and globally centrosymmetric altermagnets, in which nonrelativistic altermagnetic spin polarization and relativistic spin canting cooperate to generate a nonlinear spin-orbit torque despite the absence of any linear current-induced spin polarization.

\section{Nonlinear spin polarization}\label{Sec_NSP}

In inversion symmetry-broken materials, an applied electric field ${E}$ generates a linear spin polarization $\delta S^{(1)}$ via the Rashba-Edelstein effect (or its nonrelativistic counterpart~\cite{GonzlezHernndez2024, atasi_NC2025}), $\delta S_a^{(1)}=\alpha_{a;b}E_b$, where $a,b$ are the Cartesian coordinates and $\alpha_{a;b}$ is the linear spin susceptibility tensor. This linear response vanishes in centrosymmetric
systems. Centrosymmetric materials can, however, host a second-order nonlinear spin polarization~\cite{Xiao_prl2022,Xiao_prl2023,Sarkar2025b}, %
\be 
\delta S_a^{(2)}=\alpha_{a;bc} E_b E_c ~.
\ee
Here, $\alpha_{a;bc}$ is the nonlinear spin susceptibility tensor, which can be finite even in locally centrosymmetric materials. Under spatial inversion ($\cal P$), the spin polarization being an axial vector, remains unchanged, whereas the electric field changes sign, $E_a \rightarrow - E_a$. Consequently, the quadratic combination $E_b E_c$ is inversion-even, allowing $\alpha_{a;bc}$ to be $\mathcal{P}$-even and the second-order spin polarization to survive inversion symmetry.

We calculate the nonlinear spin polarization within a quantum kinetic framework based on the density matrix $\rho(\boldsymbol{k},t)$, which evolves according to the quantum Liouville equation,
$i\hbar\,\partial_t\rho(\boldsymbol{k},t)=[\mathcal{H},\rho(\boldsymbol{k},t)]$, with $\mathcal{H}=\mathcal{H}_0+\mathcal{H}_E$. Here, $\mathcal{H}_0$ is the Bloch Hamiltonian satisfying
$\mathcal{H}_0|u_{m\boldsymbol{k}}\rangle=\varepsilon_{m\boldsymbol{k}}|u_{m\boldsymbol{k}}\rangle$, while
$\mathcal{H}_E=e\hat{\boldsymbol r}\cdot\boldsymbol E$ describes the coupling to a uniform dc electric field $\boldsymbol E=(E_x,E_y,E_z)$. In the weak-field regime, the density matrix can be expanded perturbatively as
$\rho=\rho^{(0)}+\rho^{(1)}+\rho^{(2)}+\cdots$, with $\rho^{(N)}\propto|\boldsymbol E|^N$. The equilibrium density matrix $\rho^{(0)}$ is diagonal in the band basis, with diagonal elements given by the Fermi-Dirac distribution
$f_m^0=\left[1+\exp\left((\varepsilon_{m\boldsymbol{k}}-\mu)/k_BT\right)\right]^{-1}$. We use adiabatic switching, $\bm E(t)=\bm E e^{t/\tau}$ for $t\leq0$, with $\tau$ being the scattering timescale. The quantum Liouville equation is solved iteratively to obtain the second-order density matrix $\rho^{(2)}(\boldsymbol{k})$~\cite{sarkar_arxiv2025}.

Since our interest is in both staggered and uniform nonlinear spin responses, we evaluate the spin polarization separately on the two magnetic sublattices. The projection operator onto sublattice $\eta=A,B$ is defined as
$\hat{P}_{\eta}=\sum|\psi_\eta\rangle\langle\psi_\eta|$, where $|\psi_\eta\rangle$ denotes a localized orbital basis state belonging to sublattice $\eta$. The corresponding sublattice-resolved nonlinear spin polarization can be evaluated as
\be
\delta S^{(2),\eta}_a
=
\sum_{m,p}\int_{\bm k}
\rho^{(2)}_{mp}(\bm k)\,
s^{a,\eta}_{pm},
\ee
where
$s^{a,\eta}_{pm}
=
\frac{1}{2}
\bra{u_{p\bm k}}
(\hat{P}_{\eta}\hat{s}^a+\hat{s}^a\hat{P}_{\eta})
\ket{u_{m\bm k}}$
is the sublattice-projected spin matrix element. 
$\int_{\bm k}\equiv\int d^dk/(2\pi)^d$ denotes integration over the Brillouin zone, for spatial dimension $d$. For brevity, we henceforth write $\delta S^{(2)}_a$ simply as $\delta S_a$, since our focus is exclusively on the nonlinear spin polarization in centrosymmetric altermagnets. The uniform and N\'eel components can then be constructed from the sublattice responses as follows,
\bea\label{Uni_and_Neel}
\delta S_a^{\rm uni}=(\delta S_a^A+\delta S_a^B),\quad 
\delta S_a^{\rm N\acute{e}el}=(\delta S_a^A-\delta S_a^B).
\eea

In general, the second-order spin polarization contains several intraband and interband contributions~\cite{Sarkar2025b, Xiao_prl2022, Xiao_prl2023}. For the centrosymmetric altermagnets considered here, we focus on the nonlinear intraband Drude contribution. Its sublattice-resolved response is
\bea\label{susceps}
\alpha^{\rm D,\eta}_{a;bc} &=& -\dfrac{e^2\tau^2}{2\hbar} \sum_m\int_{\bm k} (\partial_{k_b} s_{mm}^{\eta;a})\, v_c^m \, (\partial_{\varepsilon}f_m^0)~.
\eea
Here, $v_c^m = (\partial_{k_c} \varepsilon_m)/\hbar $ is the intraband velocity. The spin response $\alpha_{a;bc}$ is, by definition, symmetric under the interchange $b\leftrightarrow c$. Note that the $\alpha^{\rm D,\eta}_{a;bc}$ is time-reversal ($\cal T$) odd response, thus, it can support staggered spin polarization between two opposite magnetic sublattices.

For completeness, a set of interband contributions to the nonlinear spin polarization is presented in Appendix~\ref{app1}. These responses can contain both $\mathcal{T}$-odd and $\mathcal{T}$-even components and arise from interband coherence and field-induced corrections to the electronic states and occupations. In the strict nonrelativistic limit, spin conservation suppresses those channels that rely on off-diagonal spin matrix elements, since $s^a_{mp}=0$ for $m\neq p$ when $[\mathcal{H}_0,{\hat s}^a]=0$~\footnote{In the strict nonrelativistic limit, the two spin sectors remain decoupled. For a conserved spin component satisfying $[\mathcal{H}_0,{\hat s}^a]=0$, the spin operator is diagonal in the eigenstate basis, such that $s^a_{mp}=0$ for $m\neq p$. If $[\mathcal{H}_0,{\hat s}^a]\neq0$, the corresponding interband spin matrix elements can in general remain finite.}. Finite SOC mixes the spin sectors and activates interband channels. Nonetheless, the nonlinear Drude term usually dominates the interband responses in metallic systems (see also Sec.~\ref {Sec_FeSb2}).

\section{Spin group symmetry analysis}

We now determine the symmetry constraints on the nonlinear spin polarization in the nonrelativistic limit. Before turning to the formal analysis, an important physical observation can be made immediately. In the absence of relativistic spin-orbit coupling (SOC), spin is conserved and the electronic states of a collinear altermagnet remain polarized along the N\'eel vector $\hat{\bm n}$. Here, {$\hat{\bm n}=(\hat{\bm m}_A-\hat{\bm m}_B)/|\hat{\bm m}_A-\hat{\bm m}_B|$} is the normalized N\'eel vector, with $\hat{\bm m}_A,\, {\hat {\bm m}}_B$ being the sublattice magnetization unit vectors. An electric field can redistribute these spin polarized itinerant electrons in momentum space, but cannot coherently rotate their spins away from the N\'eel axis. Consequently, transverse spin polarization ($\delta {\bm S} \perp {\hat{\bm n}}$) vanishes, {\it i.e.,} $\delta\bm S_{\perp}=0 $, independent of the detailed crystalline symmetry. Nonetheless, the collinear response ($\delta\bm S\parallel\hat{\bm n}$) may survive, $\delta {\bm S}_n \neq 0$. The role of spin group symmetry is to determine whether this surviving collinear nonlinear spin polarization is allowed and, crucially, whether it is uniform or staggered between the two magnetic sublattices.

In the nonrelativistic limit, the spin and lattice degrees of freedom are decoupled, and the relevant symmetry is described by spin groups rather than conventional magnetic point groups~\cite{smejkal_prx2022,Etxebarria_AC2025,Elcoro_AC2025,Elcoro_2026}. A spin group operation is written as $\{U\|R\}$, where $U$ acts in spin space and $R$ acts on the lattice coordinates. For a collinear altermagnet, the spin point group can be expressed as
\be \label{eq:SPG_decomposition}
G=G_{\rm SO}\otimes G_{\rm NT},
\ee
where $G_{\rm SO}$ is the spin-only group and $G_{\rm NT}$ is the nontrivial group. It can be further decomposed as
\be
G_{\rm NT}=\mathcal G_s+\mathcal A\mathcal G_s.
\label{eq:GNT_decomposition}
\ee
Here, $\mathcal G_s$ contains operations that preserve each magnetic sublattice, whereas the coset $\mathcal A\mathcal G_s$ contains operations that interchange the two sublattices and belong to $(G_{\rm NT} -{\cal G}_s)$. A representative sublattice-exchanging operation can be written as $\mathcal A=\{\mathcal C_2\|Q\}$, where $\mathcal C_2$ does a twofold rotation of the spin about an axis perpendicular to the collinear ordering axis $\bm{\hat n}$ and $Q$ connects the two sublattices in real space~\cite{smejkal_prx2022}.

\begin{table}[tbp]
    \caption{\textbf{Symmetry classification of nonlinear N\'eel spin polarization in the nonrelativistic limit.}
For the 10 nontrivial spin-Laue groups, the table lists the halving subgroup ${\cal G}_s$, a representative sublattice-exchanging real-space operation $Q$, the symmetry-allowed staggered collinear nonlinear spin susceptibility components, and representative material realizations. Blank material entries indicate that no representative has been established. The electric field is taken in the $xy$ plane.}
    \centering
    \renewcommand{\arraystretch}{1.75}
    \setlength\tabcolsep{0.07cm}
    \begin{tabular}{c | c | c | c | c}
    \hline
    \hline
    \centering
    {\centering $G_{\rm NT}$} & {\centering $\mathcal{G}_s$} & {\centering $Q$} & {\centering ${\alpha}_{n;bc}$} & {\centering Materials} \\
    \hline
    \multirow{2}{*}{$^2m^2m^1m$} & \multirow{2}{*}{$\{I\|2/m\}$} & \multirow{3}{*}{\centering $\mathcal{C}_{2x}$} &  \multirow{2}{*}{$\alpha^A_{n;xx}=-\alpha^B_{n;xx}$} & \ch{FeSb2},~\ch{Mn5Si3} \\
    &  &  &  \multirow{2}{*}{$\alpha^A_{n;yy}=-\alpha^B_{n;yy}$} & \ch{La2CuO4}\\
    \hhline{--~~-}
    \multirow{1}{*}{$^14/^1m^2m^2m$} & \multirow{1}{*}{$\{I\|4/m\}$} &  &  & \multirow{1}{*}{\ch{KMnF3}} \\
    \hhline{-----}
    \multirow{3}{*}{$^22/^2m$} & \multirow{3}{*}{$\{I\|\bar{1}\}$} & \multirow{3}{*}{\centering $\mathcal{C}_{2z}$} & $\alpha^A_{n;yy}=-\alpha^B_{n;yy}$ &   \multirow{3}{*}{\ch{CuF2}}\\
     &  &  & {$\alpha^A_{n;xx}=-\alpha^B_{n;xx}$} & \\
      &  &  & {$\alpha^A_{n;xy}=-\alpha^B_{n;xy}$} & \\
    \hhline{-----}
    $^24/^1m$ & $\{I\|2/m\}$ & \multirow{2}{*}{\centering $\mathcal{C}_{4z}$} & \multirow{1}{*}{$\alpha^A_{n;xx}=-\alpha^B_{n;yy}$} & \ch{KRu4O8}\\ 
    \hhline{--~~-}
    \multirow{1}{*}{$^24/^1m^2m^1m$} & \multirow{1}{*}{$\{I\|mmm\}$} &  & \multirow{1}{*}{$\alpha^A_{n;yy}=-\alpha^B_{n;xx}$} & \ch{MnF2}, \ch{MnO2}\\
    \hhline{-----}
    \multirow{1}{*}{$^16/^1m^2m^2m$} & \multirow{1}{*}{$\{I\|6/m\}$} & \multirow{2}{*}{\centering $\mathcal{C}_{21}$} & {$\alpha^A_{n;xx}=-\alpha^B_{n;xx}$} &  \\
    \hhline{--~~-}
    \multirow{1}{*}{$^1\bar{3}^2m$} & {$\{I\|\bar{3}\}$} &  & {$\alpha^A_{n;yy}=-\alpha^B_{n;yy}$} & \ch{CoF3}, \ch{Fe2O3} \\
    \hline  
    {$^26/^2m$} & {$\{I\|\bar{3}\}$} & \multirow{2}{*}{\centering $\mathcal{C}_{6z}$} & {$\alpha^A_{n;xx}=-\alpha^B_{n;xx}$} &  \\
    \hhline{--~~-}
   {$^26/^2m^2m^1m$} & $\{I\|\bar{3}m\}$ &  & {$\alpha^A_{n;yy}=-\alpha^B_{n;yy}$} & \ch{CrSb}, \ch{MnTe} \\
    \hline
    \multirow{2}{*}{$^1m^1\bar{3}^2m$} & \multirow{2}{*}{$\{I\|m\bar{3}\}$} & \multirow{2}{*}{\centering $\mathcal{C}_{4z}$} & $\alpha^A_{n;xx}=-\alpha^B_{n;xx}$ &  \\
    & & & $\alpha^A_{n;yy}=-\alpha^B_{n;yy}$ & \\
    \hline
    \hline
    \end{tabular}
    \label{tab1}
\end{table}

Since the real-space inversion $\{I\|\mathcal P\}$ belongs to $\mathcal G_s$ for all $10$ nontrivial spin-Laue groups~\cite{Smejkal_prx22_emerging}, the linear spin polarization vanishes independently on each sublattice. Therefore, the leading electrically induced spin polarization is second order, $\delta S_a = \alpha_{a;bc}E_bE_c$~\cite{Xiao_prl2022,Xiao_prl2023,Sarkar2025b}. To formulate the general symmetry constraints, we consider the sublattice-resolved susceptibility $\alpha_{a;bc}^\eta$. For a sublattice-preserving operation, the nonlinear susceptibility on each sublattice transforms as
\be
\alpha^{A/B}_{a';b'c'}
=
U_{a'a}R_{b'b}R_{c'c}
\alpha^{A/B}_{a;bc},
\label{SymRel_same}
\ee
for all $\{U\|R\}\in G_{\rm SO}\otimes \mathcal{G}_s$ [see Eq.~\eqref{eq:SPG_decomposition} and \eqref{eq:GNT_decomposition}]. In contrast, sublattice-exchanging operations relate the response on one sublattice to that on the other,
\bea
\alpha^{A/B}_{a';b'c'}
&=&
U_{a'a}R_{b'b}R_{c'c}\,
\alpha^{B/A}_{a;bc},
\label{SymRel_diff}
\eea
for all $\{U\|R\}\in G_{\rm SO}\otimes \mathcal{A}\mathcal{G}_s$. Here, summation over repeated indices is implied.

The constraint that $\delta {\bm S}_\perp=0$ for all collinear altermagnets also follows directly from $G_{\mathrm{SO}}$. For collinear magnetic order along $\hat{\bm n}$, $G_{\rm SO}$ contains arbitrary spin rotations $\{\mathcal C_{\infty{n}_\parallel}\|I\}$ about the N\'eel axis and an anti-unitary spin reversal $\{{\cal C}_{2{n_\perp}}\mathcal{T}\|\mathcal{T}\}$~\cite{smejkal_prx2022,Etxebarria_AC2025,Elcoro_2026}. Here, ${\cal C}_{\infty n_{\parallel}}$ (${\cal C}_{2 n_\perp }$) denotes arbitrary (twofold) spin rotation parallel (perpendicular) to ${\hat {\bm n}}$. While the anti-unitary operation keeps $\alpha^{\eta}_{a;bc}$ invariant, taking $\{\mathcal C_{2 n_{\parallel}}\|I\}$ as a representative operation for $\{\mathcal C_{\infty n_{\parallel}}\|I\}$, Eq.~\eqref{SymRel_same} gives
\be
\alpha^{A/B}_{a';bc}=(\mathcal C_{2n_{\parallel}})_{a'a}\alpha^{A/B}_{a;bc},
\ee
which requires $\alpha^{A/B}_{a;bc}=0$ for every spin index $a$ transverse to $\hat{\bm n}$. Thus, in the strictly nonrelativistic limit, a collinear altermagnet can generate only a collinear nonlinear spin polarization. Now, whether this response produces a net uniform spin polarization or a purely staggered N\'eel-type response is then determined by the sublattice-exchanging operations in $\mathcal A\mathcal G_s$.

Using Eqs.~\eqref{SymRel_same} and \eqref{SymRel_diff}, we classify the nonlinear spin response of all $10$ nontrivial spin-Laue groups~\cite{Elcoro_AC2025,Elcoro_2026}. The results are summarized in Table~\ref{tab1} for the applied electric field in the $xy$ plane. A striking outcome is that a \emph{uniform} nonlinear spin polarization is highly restricted. For an electric field in the $xy$ plane, symmetry permits a uniform collinear response only in the spin Laue groups $^24/^1m$, and $^24/^1m^2m^1m$. A staggered response, $\alpha_{a;bc}^A=-\alpha_{a;bc}^B$ for $a=\bm {\hat n}$, is symmetry allowed in $8$ of the $10$ spin-Laue groups. Furthermore, 6 of these 8 groups additionally allow $\alpha_{n;xx}^\eta=\alpha_{n;yy}^\eta$. Thus, even when the net electrically induced spin polarization vanishes, $\delta S_n^{\rm uni} =0$, a finite nonlinear N\'eel spin polarization, $\delta S_{n}^{\rm N\acute{e}el}$, can survive. This constitutes a purely nonrelativistic, hidden spin response intrinsic to a broad class of centrosymmetric altermagnets.

We emphasize that this nonrelativistic N\'eel spin polarization is necessarily collinear (or anti-collinear) with the local magnetic moments and does not generate a spin-orbit torque by itself. Finite SOC qualitatively changes this situation by coupling the spin and lattice degrees of freedom and reducing the relevant symmetry from the spin point group to the corresponding magnetic point group. The continuous spin-rotation symmetry about $\hat{\bm n}$ is then lost, allowing the possibility of both finite transverse nonlinear spin polarization, $\delta\bm S_{\perp}\neq0$, and new components of staggered response $\delta S_{a}^{\rm N\acute{e}el}\neq0$, with $a\neq \hat{\bm n}$~\cite{Yang_neel25}. Microscopically, SOC generates momentum-dependent canting of the itinerant electron spins away from the N\'eel axis, providing the transverse spin texture required for a torque-active nonequilibrium spin density~\cite{Brink_prb25,Rodrigo_26,Xie_26}. We demonstrate this relativistic crossover explicitly for the doped \ch{FeSb2} model~\cite{Roig_prb2024} in the following section.

\section{Nonlinear Spin Polarization in $\ch{FeSb_2}$} \label{Sec_FeSb2}

\begin{figure}[t]
    \centering
    \includegraphics[width=1 \linewidth]{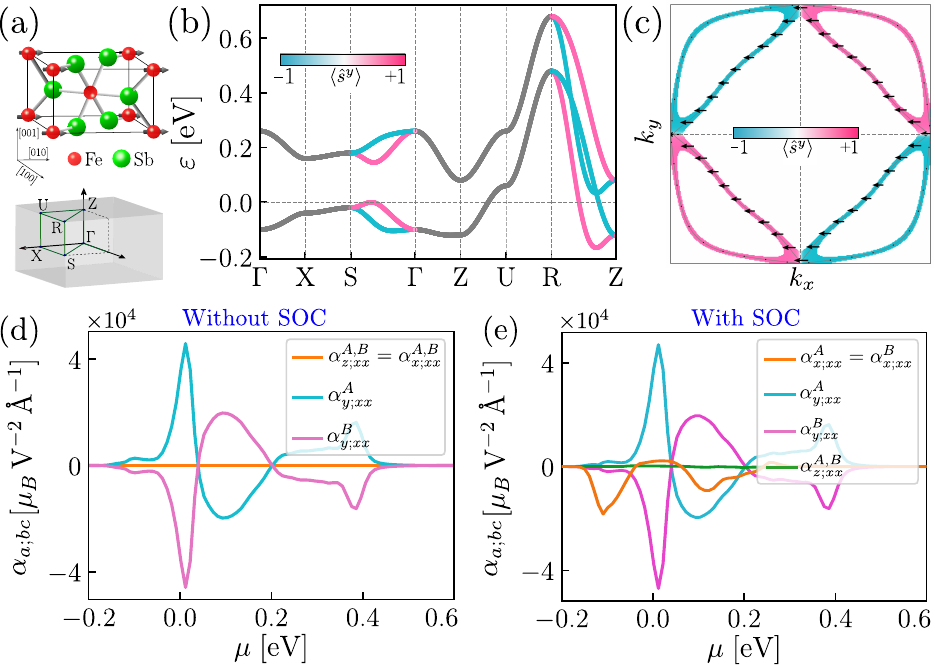}
    \caption{{\bf Nonlinear spin polarization in minimal model of doped \ch{FeSb2}.} (a) Crystal structure and Brillouin zone of \ch{FeSb2}, with the N\'eel vector oriented along $\bm {\hat n}=\hat y$. 
    (b) Spin-resolved band structure obtained from the minimal model along the $\Gamma$-$\mathrm{X}$-$\mathrm{S}$-$\Gamma$-$\mathrm{Z}$--$\mathrm{U}$-$\mathrm{R}$-$\mathrm{Z}$ high-symmetry path, exhibiting the characteristic nonrelativistic altermagnetic spin splitting along $\Gamma$-$\mathrm{S}$ and $\mathrm{Z}$-$\mathrm{R}$. The color scale denotes the spin expectation value $\langle {\hat s}^y\rangle$. 
    (c) Spin-resolved Fermi contour at $\mu=0.16~\mathrm{eV}$ in the $k_z=0$ plane. The color scale denotes $\langle {\hat s}^y\rangle$, while the arrows indicate the SOC-induced transverse spin component $\langle {\hat s}^x\rangle$. 
    (d) Sublattice-resolved nonlinear spin polarization as functions of chemical potential in the nonrelativistic limit. The collinear response is purely staggered, $\alpha^A_{y;xx}=-\alpha^B_{y;xx}$, whereas the transverse components vanish identically. 
    (e) Corresponding responses in the presence of finite relativistic SOC. The $\alpha_{y;xx}$ remains staggered, while SOC generates a finite uniform transverse response, $\alpha^A_{x;xx}=\alpha^B_{x;xx}$. In (d) and (e), we assumed $\tau=1$ ps. }
    \label{fig2}
\end{figure}

With the spin group symmetry constraints on the nonlinear spin polarization response established, we now investigate the spin polarization in doped \ch{FeSb2}. While pristine \ch{FeSb2} is a semiconductor, hole doping is predicted to stabilize metallic $d$-wave altermagnetic order~\cite{Smejkal_pnas2021}. \ch{FeSb2} crystallizes in an orthorhombic structure with its space group being $Pnnm$ [see Fig.~\ref{fig2}(a)]. The predicted hole-doped phase has its easy axis along $\hat y$~\cite{Smejkal_pnas2021}. In the nonrelativistic limit, altermagnetic \ch{FeSb2} belongs to the spin-Laue group $^2m^2m^1m$~\cite{smejkal_prx2022,Xie_26}. As summarized in Table~\ref{tab1}, this symmetry permits a nonlinear N\'eel spin polarization along the magnetic-ordering direction, $\delta S_y^A=-\delta S_y^B$, while transverse nonlinear spin polarization components are forbidden in the nonrelativistic limit.

To calculate the nonlinear spin polarization, we use a minimal tight-binding model~\cite{Roig_prb2024,Smejkal_pnas2021}, described in Appendix~\ref{app2}. The model captures the characteristic momentum-dependent spin splitting of \ch{FeSb2}, which originates from the combination of the exchange term $J_y\tau_z\sigma_y$ and the sublattice-dependent hopping $t_{z,\bm k}\tau_z$, with $t_{z,\bm k}\propto\sin k_x\sin k_y$, where $\tau_z,\,\sigma_y$ denotes the Pauli matrices corresponding to sublattice and spin spaces and $J_y$ is the exchange parameter. The crystal structure, Brillouin zone, and spin-resolved band dispersion are shown in Figs.~\ref{fig2}(a,b). The spin splitting is evident along the $\Gamma$-$\mathrm{S}$ and $\mathrm{R}$-$\mathrm{Z}$ directions, where $\sin k_x\sin k_y \neq0$.

Figure~\ref{fig2}(c) shows the corresponding spin-resolved Fermi surfaces in the $k_z=0$ plane. The characteristic $d$-wave structure of the altermagnetic spin splitting is clearly visible, with the equilibrium spin polarization remaining invariant under $\bm{k}\rightarrow-\bm{k}$. To illustrate the effect of relativistic spin-orbit coupling, the Fermi contours are calculated in the presence of finite SOC, which induces additional spin texture in the plane transverse to the N\'eel axis. In Fig.~\ref{fig2}(c), this SOC-induced itinerant spin component along $\hat x$ is indicated by the black arrows.

To analyze the spin polarization responses, we first consider the strictly nonrelativistic limit and calculate the intraband Drude contribution $\alpha_{y;xx}^{\rm D}$ as a function of the chemical potential $\mu$. As shown in Fig.~\ref{fig2}(d), the two sublattices develop nonlinear spin polarizations of equal magnitude and opposite sign, $\alpha_{y;xx}^{A}=-\alpha_{y;xx}^{B}$, over the entire range of chemical potential considered. Thus, the response is purely staggered, with $\delta S_y^{\rm N\acute{e}el}  \neq 0$, and $\delta S_y^{\rm uni} =0$ for an electric field along $\hat{x}$ axis. The spin response peaks around $\mu=0$, due to the high value of band velocity $v_x$, and $\partial s^y / \partial k_x$. Note that other nonlinear spin responses, such as $\alpha_{z;xx}$ and $\alpha_{x;xx}$, are identically vanishing [see Fig.~\ref{fig2}(d)], consistent with spin-Laue group symmetry prediction in Table~\ref{tab1}.

We next examine how this response is modified by finite spin-orbit coupling. For the \ch{FeSb2} model, we find that SOC activates a finite $\alpha_{x;xx}$ response. As shown in Fig.~\ref{fig2}(e), this transverse spin polarization is uniform between the two magnetic sublattices, $
\alpha_{x;xx}^A = \alpha_{x;xx}^B$, in sharp contrast to the staggered collinear response. At the same time, SOC has almost no effect on $\alpha_{y;xx}$. This robustness in $\alpha_{y;xx}$ reflects the nonrelativistic origin of the collinear N\'eel spin polarization, which is primarily governed by the collinear altermagnetic exchange splitting rather than by relativistic spin-orbit coupling. Furthermore, we mention that finite SOC modifies the interband contributions to the spin polarization, which has been added in $\alpha_{x;xx}$ of Fig.~\ref{fig2}(e). However, we find that interband responses are orders of magnitude smaller than the intraband Drude response.

\begin{figure}[t]
    \centering
    \includegraphics[width=1\linewidth]{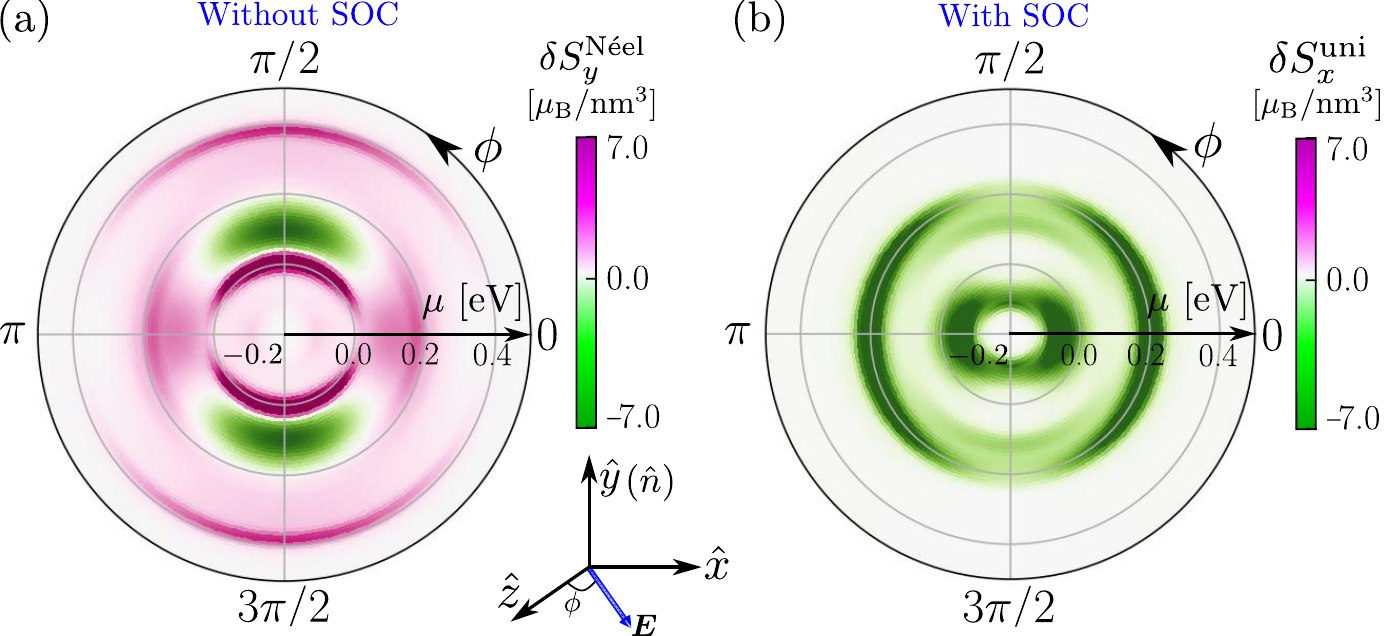}
    \caption{{\bf Angular variation of nonlinear spin polarization.} Angular dependence of (a) the nonlinear N\'eel spin polarization $\delta S_y^{\rm N\acute{e}el}$ without SOC and (b) the SOC-induced $\delta S_x^{\rm uni}$, as the electric field $\bm E$ is rotated in the $xz$ plane. The radial coordinate denotes the chemical potential $\mu$. Here, we have used $E = 3\times10^6~\mathrm{V/m}$ and $\tau=1~\mathrm{ps}$. Note that  SOC-induced changes in $\delta S_y^{\rm N\acute{e}el}$ are negligible, hence not added in (a).}
    \label{fig3}
\end{figure}

In Figs.~\ref{fig3}(a) and \ref{fig3}(b), we show $\delta S_y^{\rm N\acute{e}el}$ and the SOC-induced $\delta S_x^{\rm uni}$, respectively, as functions of $\mu$ and the orientation of the electric field in the $xz$ plane, $\phi$. The electric field can be parametrized as [see Fig.~\ref{fig3}] $\bm E=E(\sin\phi\,\hat{x}+\cos\phi\,\hat{z})$,
such that $\phi=0$ ($\pi/2$) corresponds to $\bm E\parallel\hat{z}$ ($\hat x$). In general, using Eq.~\eqref{Uni_and_Neel} the angular dependence can be written as
\bea
\delta S_a^{\rm N\acute{e}el/uni} &=&
2E^2\left[
\alpha^A_{a;xx}\sin^2\phi
+\alpha^A_{a;xz}\sin2\phi
\right. \nn \\
&& \left. +\alpha^A_{a;zz}\cos^2\phi
\right]~,
\eea
where we have used $\alpha^A_{a;bc}=-\alpha^B_{a;bc}$ for the N\'eel response and $\alpha^A_{a;bc}=\alpha^B_{a;bc}$ for the uniform response. We find that for the \ch{FeSb2} model considered here, $\alpha_{x;xz}$ and $\alpha_{y;xz}$ components vanish. Hence, the angular dependences are governed entirely by the relative magnitudes and signs of the $\alpha_{a;xx}$ and $\alpha_{a;zz}$ components, associated with $\sin^2\phi$ and $\cos^2\phi$ angular characteristics, respectively. As shown in Fig.~\ref{fig3}(a), the nonrelativistic N\'eel spin polarization $\delta S_y^{\rm N\acute{e}el}$ is strongly anisotropic: it is weaker for $\bm E\parallel\hat{ z}$, while for $\bm E\parallel\hat{ x}$ it is enhanced and reverses sign with $\mu$. In contrast, the SOC-induced uniform transverse response in Fig.~\ref{fig3}(b) retains its sign over the considered $\mu$ range and is maximal for $\bm E\parallel\hat{ z}$.

Importantly, the inclusion of SOC does not break spatial inversion symmetry in the \ch{FeSb2} model~\cite{Roig_prb2024}. Consequently, the linear current-induced spin polarization remains forbidden, and the leading electrically generated spin response continues to be second order in the applied field. The nonlinear spin polarization on each sublattice can be expressed as
\be\label{spin_sublat}
\delta\bm{S}^{A} =\frac{1}{2}(\delta {\bm S}_{n}^{\rm N\acute{e}el} +
\delta\bm{S}_{\perp}^{\rm uni}),~~
\delta\bm{S}^{B} = \frac{1}{2}(-\delta {\bm S}_{n}^{\rm N\acute{e}el} + \delta\bm{S}_{\perp}^{\rm uni}),
\ee
where $\delta {\bm S}_{n}^{\rm N\acute{e}el}$ denotes the nonrelativistic staggered component and $\delta\bm{S}_{\perp}^{\rm uni}$ the SOC-induced uniform transverse component. Consequently, even in centrosymmetric altermagnets, a sublattice-wise asymmetric nonlinear spin polarization is developed, {\it i.e.},  $\delta\bm{S}^{A}\neq-\delta\bm{S}^{B}$.

This sublattice-asymmetric spin response has an important consequence for the N\'eel order dynamics. The N\'eel spin polarization $\delta\bm S_{n}^{\rm N\acute{e}el}$ is collinear with the local magnetic moments and therefore it does not generate a spin-orbit torque by itself. In contrast, the SOC-induced transverse component $\delta\bm S_{\perp}^{\rm uni}$ is torque-active and couples the electrically generated spin polarization to the N\'eel order dynamics. However, $\delta\bm S_{\perp}^{\rm uni}$ alone can not switch the N\'eel order, without finite $\delta\bm S_{n}^{\rm N\acute{e}el}$. In the following section, we show how the coexistence of these components gives rise to the torques required for deterministic, all-electrical switching of the N\'eel order in centrosymmetric altermagnets.

\section{N\'eel Order Switching in $\ch{FeSb2}$}

Having established the sublattice-resolved nonlinear spin response in \ch{FeSb2}, we now examine whether it can induce deterministic switching of the N\'eel order. We employ a two-sublattice macro-spin description and solve the coupled Landau-Lifshitz-Gilbert (LLG) equations for the unit magnetization vectors $\hat{\bm m}_{A/B}$,
\be
\frac{d\hat{\bm m}_{\eta}}{dt} = -\gamma \hat{\bm m}_{\eta}\times\bm B^{\rm eff}_{\eta} +
\alpha_G\hat{\bm m}_{\eta}\times\frac{d\hat{\bm m}_{\eta}}{dt} + \bm\tau^{\rm FL}_{\eta} + \bm\tau^{\rm DL}_{\eta}~. \label{Dynamic_AFM}
\ee
Here, $\gamma$ is the gyromagnetic ratio and $\alpha_G$ is the Gilbert damping parameter. The effective field $\bm B_\eta^{\rm eff}$ contains the intersublattice exchange and magnetic-anisotropy contributions,
\be
\bm B^{\rm eff}_{A/B}=-B_E \hat{\bm m}_{B/A}+ B_K(\hat{\bm m}_{A/B}\cdot\hat{\bm e})\hat{\bm e} + \bm B_{\rm ext}~.
\ee
Here, ${\bm B}_E$ is the exchange field, and ${\bm B}_{\rm ext}$ is the applied external magnetic field, which we assume to be zero. $\hat{\bm e}$ denotes the magnetic easy axis and ${\bm B}_K$ is the anisotropy field. The current-induced field-like (FL) and damping-like (DL) spin torques are generated by the corresponding sublattice-resolved nonequilibrium spin polarization and take the conventional forms
\be
\begin{aligned}
\bm\tau_\eta^{\rm FL} &=|{\bm \tau}_q|\xi^{\rm FL}
\hat{\bm m}_\eta\times\delta\hat{\bm S}^\eta, \\
\bm\tau_\eta^{\rm DL} &=|{\bm \tau}_q|\xi^{\rm DL} \hat{\bm m}_\eta\times \left( \hat{\bm m}_\eta\times\delta\hat{\bm S}^\eta \right).
\end{aligned}\label{spin_torques}
\ee
The characteristic torque magnitude is
$|{\bm \tau}_q| \approx {\gamma B_E |\delta S_{a}^\eta|}/{M_s}$~\cite{Cogulu_prl2022,Kondou_NC2021,Zhang_prb2015,Xu_JAP2023}. Here, $M_s$ is the sublattice saturation magnetization and $|\delta S_a^\eta|$ denotes the magnitude of the current-induced spin polarization component. The dimensionless parameters $\xi^{\rm FL}$ and $\xi^{\rm DL}$ characterize the FL and DL torque efficiencies, respectively, and depend on material parameters and device geometry~\cite{Sethu_pra2021,Amin_prb2016,Garello_nature2013,Manchon_RMP2019,Xu_JAP2023}.

For an applied electric field $\bm E=3\times10^6~\hat{x}$ ${\rm V/m}$ and chemical potential $\mu=0.1~{\rm eV}$, the nonlinear spin response obtained from Fig.~\ref{fig2} is $\delta\bm S^{A/B} = (-0.72\,\hat{x} \pm 1.80\,\hat{y})$ $\rm  \mu_B/{\rm nm}^{3}$. Thus, the nonlinear response consists of a uniform transverse spin polarization along $\hat{x}$ and a staggered collinear spin polarization along $\hat{y}$. For the macro-spin simulations, we take the sublattice magnetization to be $M_s\simeq1.5\times10^{5}~{\rm A/m}$, an intersublattice exchange field $B_E\simeq190~{\rm T}$, an easy-axis anisotropy field $B_K\simeq3.3~{\rm T}$, and a damping parameter $\alpha_G=1.85\times10^{-2}$. These values specify the phenomenological macro-spin model. We further take $\xi^{\rm DL}/\xi^{\rm FL}=0.1$ for the relative strength of the DL and FL torques.

\begin{figure}[t]
    \centering
    \includegraphics[width=1\linewidth]{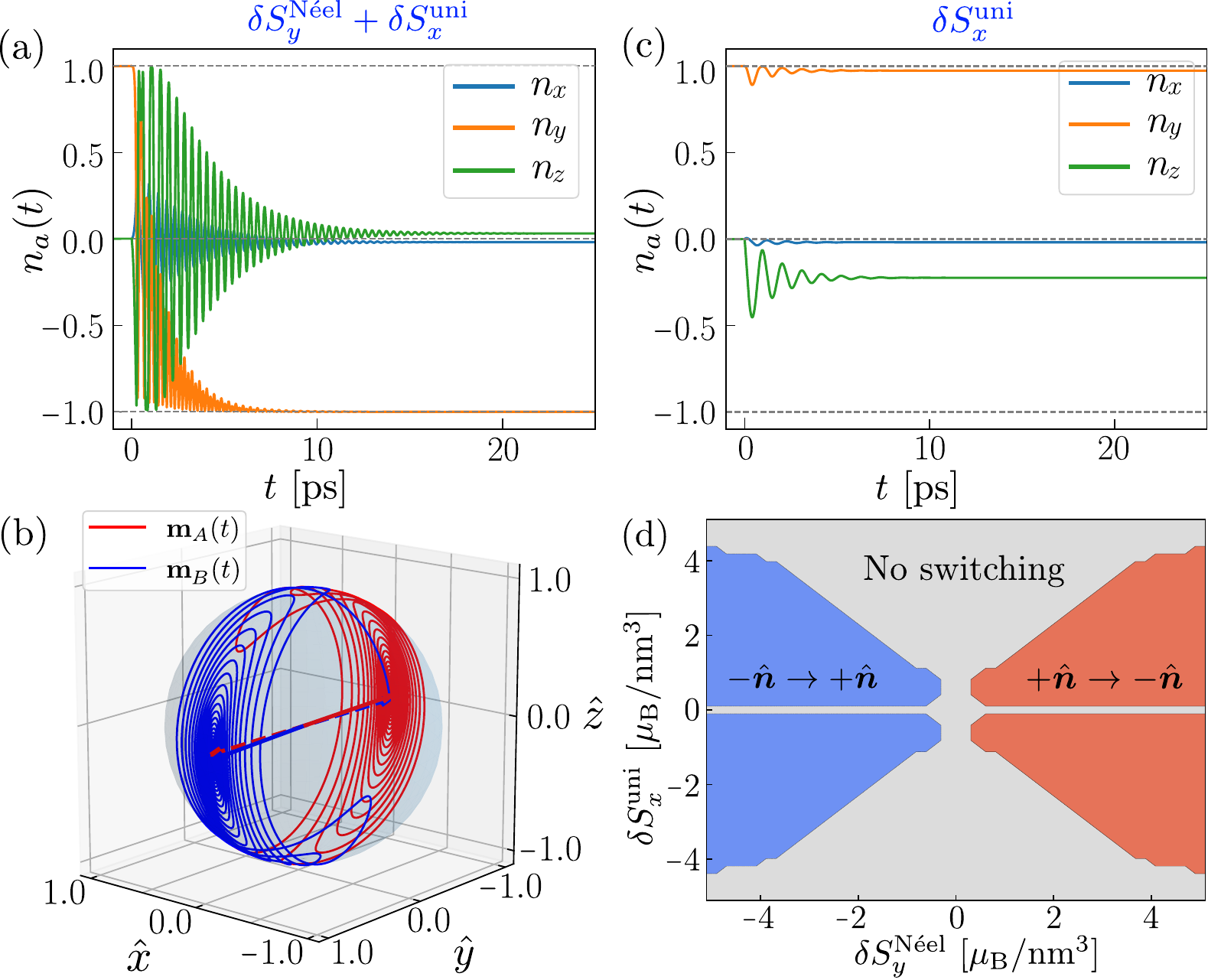}
    \caption{\textbf{N\'eel order dynamics of \ch{FeSb2}.}
    (a) Time evolution of the N\'eel vector components under the combined action of spin-orbit torque generated from the $\delta S^{\rm N\acute{e}el}_y$ and $\delta S^{\rm uni}_x$ showing deterministic $\sim180^\circ$ switching. Here, we used the values of $\delta S^{\rm N\acute{e}el}_y$ and $\delta S^{\rm uni}_x$ for $\mu = 0.1$ eV, and $\bm{E}=3\times 10^6\, \hat{x} ~\mathrm{V/m} $. (b) Three-dimensional trajectories of the two sublattice magnetizations during the switching process. (c) N\'eel order dynamics in the presence of only the SOC-induced transverse spin polarization, exhibiting damped oscillations toward a tilted state without deterministic switching.
    (d) Switching phase diagram as a function of the staggered collinear and transverse spin polarization amplitudes. Blue and brown regions denote successful switching into the reversed configurations, while the gray region corresponds to incomplete switching or oscillations.}
    \label{fig4}
\end{figure}

The resulting local magnetization dynamics are presented in Fig.~\ref{fig4}. Figure~\ref{fig4}(a) shows the time evolution of the normalized N\'eel vector $\hat {\bm n}$. Starting from $\hat {\bm n}\parallel+\hat{ y}$, the N\'eel vector reverses to $\hat{\bm  n}\parallel-\hat{ y}$ within approximately $15~{\rm ps}$, demonstrating ultrafast $180^\circ$ switching. The corresponding trajectories of the two sublattice moments are shown in Fig.~\ref{fig4}(b). The switching proceeds through a transient canting of the otherwise nearly antiparallel magnetic moments. Owing to the large exchange field, even a small canting generates a strong exchange torque, driving the rapid precessional dynamics of the N\'eel vector toward the reversed configuration.

The two components of the nonlinear spin polarization play distinct but complementary roles in the switching dynamics. Initially, the staggered collinear component is collinear with the corresponding local magnetic moments and therefore it cannot initiate switching by itself. In contrast, the uniform transverse component is noncollinear with both sublattice moments and generates the finite spin torque required to initiate their canting. This canting activates the much larger intersublattice exchange torque and thereby drives the rapid rotation of the N\'eel vector. Importantly, however, the transverse component alone is insufficient to produce deterministic reversal. As shown in Fig.~\ref{fig4}(c), retaining only the transverse uniform spin polarization leads to damped oscillations toward a canted steady state, rather than complete N\'eel order reversal~\footnote{This behavior is consistent with the understanding that a uniform sublattice spin polarization alone usually does not produce deterministic switching and can instead drive oscillatory dynamics in antiferromagnets~\cite{Shao_25}.}. Within the present macro-spin model, stable $180^\circ$ reversal results from the cooperative dynamics of the staggered collinear and
uniform transverse spin responses, assisted by the strong intersublattice exchange, rather than from either a purely uniform torque or an amplitude imbalance between otherwise collinear sublattice spin accumulations.

Figure~\ref{fig4}(d) presents the switching phase diagram as a function of the staggered collinear and uniform transverse spin polarization amplitudes. The two switched regions (blue and brown) correspond to relaxation toward the two opposite easy-axis configurations, whereas the intermediate gray region represents incomplete reversal or oscillatory dynamics~\footnote{The phase-space diagram of Fig.~\ref{fig4}(d) assumes the cut-off angle of N\'eel vector rotation for a complete switching to be $175\degree$.}. The finite extent of the switched regions demonstrates deterministic N\'eel order reversal over a broad range of torque amplitudes. More importantly, Fig.~\ref{fig4}(d) shows that even a small transverse spin polarization component can enable switching in the presence of a finite N\'eel spin polarization.

\begin{figure}[t]
    \centering
    \includegraphics[width=1\linewidth]{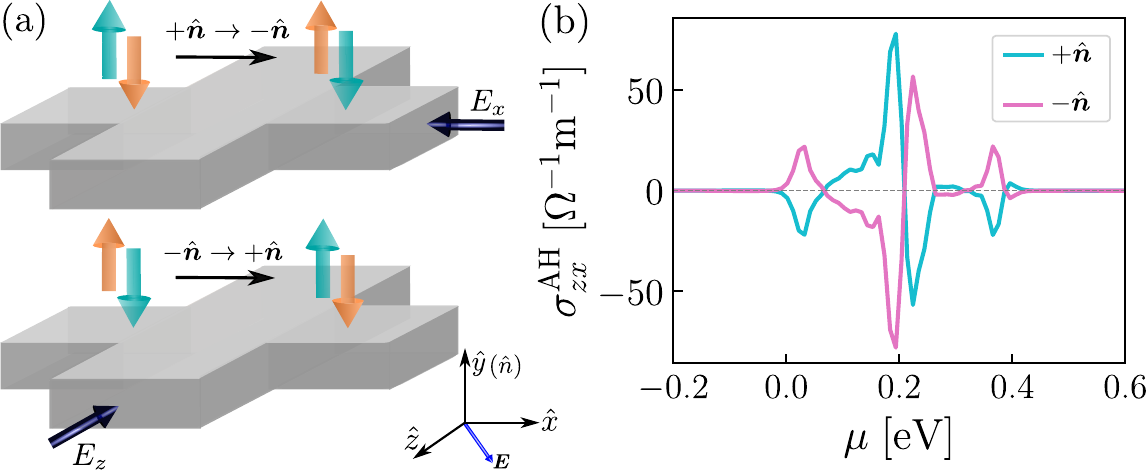}
    \caption{\textbf{Electrical write-read scheme for the N\'eel order in \ch{FeSb2}.}
    (a) Schematic device configurations for deterministic switching between the two opposite N\'eel states, achieved by changing the orientation of the applied electric field. The upper (lower) configuration illustrates the electric-field geometry used for switching between the corresponding opposite N\'eel orientations.
    (b) Anomalous Hall conductivity $\sigma_{zx}^{\rm AH}$ as a function of chemical potential for the two N\'eel configurations, $+\hat{\bm n}$ and $-\hat{\bm n}$. Reversal of the N\'eel vector reverses the sign of the anomalous Hall response, providing an electrical readout of the magnetic state.}
    \label{fig5}
\end{figure}

Furthermore, the phase diagram also shows that switching between the two opposite N\'eel states, $+\hat {\bm n} \rightarrow - \hat {\bm n}$ and $-\hat {\bm n}\rightarrow + \hat {\bm n}$, requires a corresponding reversal of the staggered nonequilibrium spin polarization. Since this response is quadratic in the electric field, reversing $\bm E$ alone does not reverse its sign. Instead, its sign can be controlled by changing the electric field orientation, or possibly chemical potential. As demonstrated in Fig.~\ref{fig3}(a), rotating $\bm E$ within the $xz$ plane can reverse the staggered nonlinear spin polarization in \ch{FeSb2}. Based on this angular dependence, Fig.~\ref{fig5}(a) illustrates a possible device configuration and the corresponding electric field orientations for realizing both $+\hat {\bm n}\rightarrow-\hat {\bm n}$ and $- \hat {\bm n} \rightarrow + \hat {\bm n}$ switching states, required for writing processes in altermagnetic memory devices~\footnote{The specific electric field orientations illustrated in Fig.~\ref{fig5}(a) follow from the angular dependence obtained within the present minimal \ch{FeSb2} model. The optimal field orientations and switching thresholds may vary with the detailed material-specific electronic structure.}.

Beyond electrical writing, a practical altermagnetic memory also requires electrical readout of the final N\'eel state. Since the two opposite configurations, $+\hat{\bm n}$ and $-\hat{\bm n}$, are related by time reversal, their anomalous Hall conductivities should reverse sign. For the \ch{FeSb2} model considered here, we find $\sigma_{zx}^{\rm AH}(+\hat{\bm n})=-\sigma_{zx}^{\rm AH}(-\hat{\bm n})$, as shown in Fig.~\ref{fig5}(b). A finite allowed Hall response can provide a direct electrical readout of the switched N\'eel configuration, completing the write-read scheme in which the N\'eel order is written by nonlinear spin polarization-driven switching and read out through its Hall response. Having established the switching mechanism and its electrical write-read functionality in \ch{FeSb2}, we now place these results in the broader context of symmetry-allowed nonlinear spin responses in centrosymmetric altermagnets.

\section{Discussion}

The spin group classification in Table~\ref{tab1} applies to the strict nonrelativistic limit, where the nonlinear spin polarization is constrained to the N\'eel axis and a staggered collinear response is allowed in 8 of the 10 nontrivial spin-Laue groups. Finite SOC reduces the relevant symmetry to the magnetic point group and can activate additional transverse components, whose uniform or staggered character is determined by the relativistic crystal symmetry~\cite{Yang_neel25}.

The contrast between \ch{FeSb2} and \ch{CrSb} illustrates these two possibilities. In \ch{FeSb2}, the nonrelativistic staggered response $\delta S_y^{A}=-\delta S_y^{B}$ coexists, in the presence of SOC, with a uniform transverse component, $\delta S_x^{A}=\delta S_x^{B}$, giving the cooperative switching mechanism demonstrated above. In \ch{CrSb}, the nonrelativistic ${}^{2}6/{}^{2}m{}^{2}m{}^{1}m$ spin-Laue group likewise allows a staggered collinear response. With SOC, however, its magnetic point group $6'/m'mm'$ permits additional staggered transverse responses, including $\alpha^{A}_{x;xz}=-\alpha^{B}_{x;xz}$ and $\alpha^{A}_{y;yz}=-\alpha^{B}_{y;yz}$, which directly generate a nonlinear N\'eel torque~\cite{Yang_neel25}. Thus, the nonrelativistic spin group response yields the parent collinear spin polarization, while SOC can generate qualitatively different transverse and collinear channels, depending on the magnetic point group.

Finally, surface and two-dimensional altermagnetism~\cite{Colin_prx26, Rodrigo_26, Jana_prb25, Jana2026, Sun_prl26} provide another natural extension of the present mechanism. At a surface or interface, the reduced crystal symmetry can relax the bulk constraints on the spin response and permit additional spin-orbit coupled terms, including Rashba-like contributions when allowed by the surface point group, thereby possibly activating spin polarization components that are forbidden in the corresponding bulk. Nonlinear spin polarization responses at surface altermagnets, and altermagnetic interfaces have already been predicted~\cite{Brink_prb25, Rodrigo_26}, while recent works demonstrate that surface symmetry breaking can itself generate or modify altermagnetic spin splitting~\cite{Colin_prx26}. This suggests a particularly interesting possibility: a bulk response that is purely collinear and torque-inactive may acquire an additional torque-active surface component, providing a route to electrically manipulate the N\'eel order through surface or interface engineering. 

\section{Conclusion}

In summary, we have established a nonlinear route to electrical control of the N\'eel order in centrosymmetric altermagnets, where inversion symmetry forbids linear current-induced spin polarization. In the nonrelativistic limit, spin group symmetry constrains the nonlinear spin polarization to lie along the N\'eel vector, with a staggered N\'eel response allowed in $8$ of the $10$ nontrivial spin-Laue groups. Although this collinear component is torque-inactive by itself, finite spin-orbit coupling can generate an additional transverse nonlinear spin polarization. Using doped \ch{FeSb2} as a representative $d$-wave altermagnet, we demonstrate the coexistence of the nonrelativistic staggered collinear and SOC-induced uniform transverse spin responses.

The two spin polarization components play complementary roles in the magnetization dynamics: the transverse component initiates canting of the sublattice moments, while the staggered component, together with the strong antiferromagnetic exchange, enables stable $180^\circ$ reversal. Our macro-spin LLG simulations show deterministic switching over a finite range of spin polarization amplitudes, while the transverse component alone produces incomplete switching or oscillations. Furthermore, the anisotropic nonlinear response allows the sign of the staggered spin polarization, and hence the switching direction, to be controlled by the orientation of the applied electric field. These results establish nonlinear spin polarization as a route to deterministic, all-electrical switching of the N\'eel order in centrosymmetric altermagnets without requiring inversion-symmetry breaking.

\section*{Acknowledgments}
S.D. is supported by the Fellowship for Academic and Research Excellence (FARE) program, IIT Kanpur.  S.S. acknowledges IIT Kanpur for funding support. A.A. acknowledges funding from the Core Research Grant by ANRF (Sanction No. CRG/2023/007003), Department of Science and Technology, India.

\section*{Data Availability}
The data that support the findings of this article are not
publicly available. The data are available from the authors
upon reasonable request.

\appendix

\section{Inter-band components of nonlinear spin susceptibility}\label{app1}

In Sec.~\ref{Sec_NSP}, we focused primarily on the intraband contribution and only mentioned the existence of interband contributions to the nonlinear spin susceptibility. Here, we discuss their microscopic origin and symmetry properties in more detail, including both interband-coherence effects and interband-induced changes in band occupations. The coherence contribution is
\be
\sum_{m\neq p}\int_{\bm k}
\rho^{(2)}_{mp}(\bm k)s^{a,\eta}_{pm}.
\ee
Together with the interband-induced diagonal terms $\sum_m\int_{\bm k}\rho_{mm}^{(2),\mathrm{inter}}s^{a,\eta}_{mm}$, the nonlinear spin susceptibility contains the following contributions
\be\begin{aligned} 
\alpha^{\eta, \rm BCP}_{a;bc}&={2e^2\hbar^2}~{\rm Re}\sum_{m\neq p}\int_{\bm k}\dfrac{s^{a,\eta}_{mm} v^b_{mp} v^c_{pm}}{(\varepsilon_m-\varepsilon_p)^3}\dfrac{\partial f_m^{0}}{\partial \varepsilon_m}~, \\ \alpha^{\eta, \rm ASM1}_{a;bc}&={2e^2\hbar}~{\rm Re}\sum_{m\neq p}\int_{\bm k}\dfrac{s^{a,\eta}_{pm} v^b_{mp}}{(\varepsilon_m-\varepsilon_p)^3}\partial_c f_m^0~, \\ \alpha^{\eta, \rm ASM2}_{a;bc}&=-{2\tau e^2}~{\rm Im}\sum_{m\neq p}\int_{\bm k}\dfrac{s^{a,\eta}_{pm} v^b_{mp}}{(\varepsilon_m-\varepsilon_p)^2}\partial_c f_m^0~,\\ \alpha^{\eta, \rm VI}_{a;bc}&={2e^2\hbar^2}~{\rm Re}\sum_{m\neq p}\int_{\bm k}\dfrac{s^{a,\eta}_{pm} (v^b_{mm}-v^b_{pp}) v^c_{mp}}{(\varepsilon_m-\varepsilon_p)^4}f_m^0~,\\ \alpha^{\eta, \rm SI}_{a;bc}&={e^2\hbar^2}~{\rm Re}\sum_{m\neq p}\int_{\bm k}
\dfrac{ (s^{a,\eta}_{pp}-s^{a,\eta}_{mm}) v^{b}_{pm} v^c_{mp}}{(\varepsilon_m-\varepsilon_p)^4}f_m^0~,\nn\\
\alpha^{\eta, \rm Sh}_{a;bc}&=2e^2~{\rm Im}\sum_{m\neq p}\int_{\bm k}\dfrac{s^{a}_{pm} {D}_{mp}^b \mathcal{R}_{mp}^c}{(\varepsilon_m-\varepsilon_{p})^2}f^0_m~.
\end{aligned}\ee
Here, $D^b_{mp}=[\partial_b-i(\mathcal{R}^b_{mm}-\mathcal{R}^b_{pp})]$ is the covariant derivative. The different contributions correspond to Berry connection polarizability (BCP), anomalous spin magnetization (ASM), velocity injection (VI), spin injection (SI) and shift (Sh) contributions.  These contributions are discussed in details in Refs.~\cite{Sarkar2025b,Xiao_prl2022,Xiao_prl2023}. Microscopically, the different terms originate from distinct field-induced corrections to the band distribution and eigenstates. The $\alpha^{\eta,\rm BCP}_{a;bc}$ term arises from the second-order correction to the distribution function associated with the field-induced energy shift. In contrast, the $\alpha^{\eta,\rm ASM1}_{a;bc}$, $\alpha^{\eta,\rm VI}_{a;bc}$, $\alpha^{\eta,\rm SI}_{a;bc}$ and $\alpha^{\eta,\rm Sh}_{a;bc}$ terms originate from the second-order correction to the band eigenstates in the presence of the electric-field perturbation. The $\alpha^{\eta,\rm ASM2}_{a;bc}$ contribution has a mixed origin, involving the first-order corrections to both the eigenstates and the distribution function.

Among these contributions, $\alpha^{\eta,\rm ASM2}_{a;bc}$ is $\mathcal{T}$-even, whereas the remaining terms are $\mathcal{T}$-odd and change sign upon reversal of the magnetic order. {In the nonrelativistic limit, the spin sectors are decoupled, and the off-diagonal spin matrix elements (satisfying $[\mathcal{H}_0,{\hat s}^a]=0$), connecting states of different spin character vanish. Consequently, the interband contributions that rely on such matrix elements are suppressed in the absence of SOC. Finite SOC mixes the spin sectors and generates nonzero off-diagonal spin matrix elements, thereby activating these interband spin polarization channels. These terms provide relativistic corrections to the dominant nonrelativistic response discussed in the main text.}

\section{Minimal Model of \ch{FeSb2}}\label{app2}

We employ a minimal tight-binding model that captures the essential symmetry and electronic structure of $d$-wave altermagnets. We adopt the model parameters of Ref.~\cite{Roig_prb2024} for \ch{FeSb2}, which crystallizes in the centrosymmetric space group $Pnnm$ and exhibits $d$-wave altermagnetic spin splitting. The Hamiltonian is given by
\be\label{Ham_AM}
{\cal H} = \varepsilon_{0,{\bm k}} + t_{x,{\bm k}} \tau_x + t_{z,{\bm k}} \tau_z + \tau_y {\bm \lambda}_{{\bm k}} \cdot \bm{\sigma} + \tau_z {\bm J} \cdot {\bm \sigma}~,
\ee
where $\tau_i$ and $\sigma_i$ are the Pauli matrices in the sublattice and spin spaces, respectively. Here, $\varepsilon_{0,{\bm k}}$ describes the sublattice-independent dispersion, while $t_{x,{\bm k}}$ and $t_{z,{\bm k}}$ represent inter- and intra-sublattice hopping, respectively. The vector $\bm{\lambda}_{\bm k}$ describes the spin-orbit coupling, and $\bm{J}$ denotes the N\'eel order parameter. For the $Pnnm$ space group, the momentum-dependent coefficients are
\bea \varepsilon_{0,\bm k}&=&t_{1x}\cos{k_x}+t_{1y}\cos{k_y}+t_2\cos{k_z}\nn\\&&+t_3\cos{k_x}\cos{k_y}+ t_{4x}\cos{k_x}\cos{k_z}\nn\\&&+t_{4y}\cos{k_y}\cos{k_z}+t_5\cos{k_x}\cos{k_y}\cos{k_z} - \varepsilon_F~,\nn \\ 
t_{x,\bm k}&=& t_8\cos(k_x/2)\cos(k_y/2)\cos(k_z/2)~,\nn \\ t_{z,\bm k} &=& t_6 \sin{k_x}\sin{k_y} + t_7\sin{k_x}\sin{k_y}\cos{k_z}~, \nn \\ \lambda_{x, \bm k}&=&\lambda_{x0} \sin(k_x/2)\cos(k_y/2)\sin(k_z/2)~, \nn\\ \lambda_{y, \bm k}&=&\lambda_{y0} \cos(k_x/2)\sin(k_y/2)\sin(k_z/2) ~, \nn \\ \lambda_{z, \bm k}&=& \lambda_{z0}\cos(k_x/2)\cos(k_y/2)\cos(k_z/2)~. 
\eea
For \ch{FeSb2} with the N\'eel vector oriented along the $y$-axis~\cite{Smejkal_pnas2021,Miller_prb03}, we use
$t_{1x}=-0.1$, $t_{1y}=-0.05$, $t_2=-0.05$, $t_3=0.06$, $t_{4x}=0.1$, $t_{4y}=0.05$, $t_5=-0.05$, $t_6=0.05$, $t_7=-0.1$, $t_8=0.15$, and $J_y=0.1$ and $\varepsilon_F = -0.12$~\cite{Roig_prb2024}, with all energies given in eV. In the absence of SOC, $\bm{\lambda}_{\bm k}=0$, the band dispersion becomes
\be
\varepsilon_{\pm} = \varepsilon_{0,{\bm k}} \pm \sqrt{t_{x, {\bm k}}^2 + (t_{z,{\bm k}} + J_y s_y)^2}~,
\ee
where $s_y=\pm1$ denotes the eigenvalue of $\sigma_y$. The spin-dependent term arises from the interplay between the intra-sublattice hopping $t_{z,\bm k}$ and the N\'eel order $J_y$. Since $t_{z,\bm k}\propto\sin k_x\sin k_y$, this coupling produces the characteristic $d$-wave form of the altermagnetic spin splitting. In this nonrelativistic limit, the nonlinear spin polarization is constrained to the N\'eel axis and is staggered between the two sublattices, consistent with the symmetry analysis in the main text [Fig.~\ref{fig2}(d)].

To examine the effect of SOC, we set $\lambda_{x0}=\lambda_{y0}=\lambda_{z0}=5$~meV. SOC breaks the continuous spin-rotation symmetry about the N\'eel axis and allows a transverse nonlinear spin polarization. As shown in Fig.~\ref{fig2}(e), a finite uniform spin response perpendicular to the N\'eel vector emerges, while the collinear staggered response remains finite. These results demonstrate the coexistence of the nonrelativistic staggered and SOC-induced transverse nonlinear spin polarizations in \ch{FeSb2} explicitly.

\bibliography{refs}
\end{document}